\documentclass[12pt]{article}
\usepackage{epsf}
\title{ {\bf
Top quark electric and chromo electric dipole moments in the general two 
Higgs Doublet model.}}
\author{\vspace{1cm}\\
        {\bf E. O. Iltan}
        \thanks{E-mail address:
        eiltan@heraklit.physics.metu.edu.tr}
 \\
        Physics Department, Middle East Technical University \\
        Ankara, Turkey\\}

\date{}

\begin{document}
\setlength{\baselineskip}{24pt}
\maketitle
\setlength{\baselineskip}{7mm}
\begin{abstract}
We study the electric and chromo electric dipole moment of top quark in the 
general two Higgs Doublet model (model III). We analyse the dependency of 
this quantity  to the new phases coming from the complex Yukawa couplings  
and masses of charged and neutral Higgs bosons. We observe that the electric 
and chromo elecric dipole moments of top quark are at the order of 
$10^{-21}\, e\,cm$ and $10^{-20}\, g_s\,cm$, which are extremely large 
values compared to the ones calculated in the SM and also the two Higgs 
Doublet model with real Yukawa couplings.
\end{abstract} 
\thispagestyle{empty}
\newpage
\setcounter{page}{1}
\section{Introduction}
Among all fermions, included in the standard model (SM), the top quark
reaches great interest, since it breaks the $SU(2)\times U(1)$ symmetry 
maximally due to its large mass. The investigation of top quark properties, 
such as mass, decay width, decay products and some fundamental quantities 
like electric dipole moment (EDM), chromo electric dipole moment (CEDM),... 
etc., becomes important. 

EDM (CEDM) of a fermion is induced by the CP violating interaction and it 
provides comprehensive informations in the determination of the free 
parameters of the various theoretical models. There are large number of 
experimental studies in the literature about EDMs of fermions. Neutron EDM 
has a special interest and the experimental upper bound has been found as 
$d_N < 1.1\times 10^{-25} e\, cm$ \cite{smith}. The electron, muon and tau 
EDMs have been measured experimentally as $d_e =(1.8\pm 1.2\pm 1.0)\times 
10^{-27} e\, cm$ \cite{Commins}, $d_{\mu} =(3.7\pm 3.4)\times 10^{-19} e\, 
cm$ \cite{Bailey} and $d_{\tau} =(3.1)\times 10^{-16} e\, cm$ \cite{Groom} 
respectively. With these measurements, it would be possible to get powerful 
clues about the internal structure of the particles, if it exists.  

The complex Cabbibo-Kobayashi-Maskawa (CKM) matrix elements are the sources 
of CP violation in the SM and therefore nonzero EDM or CEDM. However, EDM 
(CEDM) of quarks in the SM exists at the three loop level \cite{Khiplovich} 
and estimated as $\sim 10^{-30}\, (e\,(g_s)\, cm)$ even for top quark case.
Notice that quark EDM vanishes at one loop order in the SM, since moduli of 
matrix element is involved in the relevant expression. Furthermore, it also 
vanishes at two loop order after sum over internal flavours 
\cite{sahab1,donog}. When QCD corrections are taken into account, nonzero EDM 
exists \cite{krause}. Therefore, one has an open window to go beyond the SM 
and test a new physics. There are many sources of CP violation in the models 
beyond the SM, such as multi Higgs doublet models (MHDM), supersymmetric model 
(SUSY), extra dimensions \cite{Schmidt},..., etc. 

The quark EDM was calculated in the multi Higgs doublet models in the
literature \cite{Weinberg, Branco, Smith,Atwood1}. EDM, induced by the neutral Higgs 
boson effects, was studied in the two Higgs doublet model (2HDM)
\cite{Branco} and in \cite{Atwood1}, the necessity of more scalar fields than 
just two Higgs doublets was emphasised for non-zero EDM when only the charged 
Higgs boson effects were taken into account in \cite{Smith}. The EDM and 
weak EDM were calculated in the 2HDM and models with three and more Higgs 
doublets. Weak EDM of b-quark was predicted in the range 
$10^{-21}-10^{-20} e\, cm$, following the scenario where CP violation 
may only come from the neutral Higgs sector. Furthermore, $b$-quark EDM was 
obtained in the range $10^{-23}-10^{-22} e-cm$ when the CP violating effect 
comes from the charged sector. In \cite{liao}, quark EDM was calculated 
in the 2HDM if the CP violating effects are due to the  CKM matrix elements 
and it was thought that $H^{\pm}$ particles also mediate CP violation besides 
$W^{\pm}$ bosons, however, at the two loop order these new contributions 
vanish. In \cite{Dumm}, the electric and weak electric dipole form factors 
of heavy fermions in a general two Higgs doublet model were studied  and it 
was concluded that the enhancement of three orders of magnitude in the 
electric dipole form factor of the $b$ quark with respect to the prediction 
of 2HDM I and II was possible. The EDM of $b$-quark in the general 2HDM 
(model III) \cite{WSHou} and the general 3HDM with $O(2)$ symmetry in the 
Higgs sector ($3HDM(O_2)$) has been studied in \cite{eril3} and it was 
observed that a large EDM, at the order of $10^{-20}\, e\, cm$, could be 
obtained, using the complex Yukawa couplings. In \cite{Xu}, the leading 
contribution to the EDM and CEDM of the top quark was calculated in 
Higgs-boson-exchange models of CP nonconservation. The dipole moments were 
estimated of the order $10^{-20}\, (e (g_s)\, cm)$. In this work, EDM (CEDM) 
was assumed to arise at one loop order through neutral Higgs boson exchange.

In our work, we study EDM and CEDM of top quark in the model III, including 
charged Higgs contribution. In this case the sources of CP violation
are the complex Yukawa couplings $\xi^U_{N,tt}$ and $\xi^D_{N,bb}$  which 
bring two independent CP violating parameters, $sin\,\theta_{tt}$ and 
$sin\,\theta_{tb}$ (see section two for their definitions). These parameters 
play the important role for the CP violating interactions, which are
responsible for the EDM (CEDM) of top quark. It is interesting to study the 
sensitivity of top quark EDM (CEDM) to these parameters, since it gives a 
compherensive information about the new physics beyond the SM and also the 
sign of the Wilson coefficient $C_{7}^{eff}$ (see Discussion). Furthermore, 
the dependencies of EDM (CEDM) to the new Higgs boson masses, namely 
$m_{H^{\pm}}$,  $m_{h^0}$ and $m_{A^0}$, are informative in the determination 
of model parameters. This work is devoted to above analysis and presents the 
upper and lower limits of EDM (CEDM), which are obtained by using the present
experimental results. The numerical values of EDM (CEDM) of top quark can 
be estimated at the order of the magnitude of $10^{-21}\, (e \, cm)$ 
($10^{-21}-10^{-20}\, (g_s \, cm)$).      
  
The paper is organized as follows:
In Section 2, we present EDM and CEDM of top quark in the framework of model 
III. Section 3 is devoted to discussion and our conclusions.

\section{Electric and chromo electric dipole moments of top quark in the 
general two  Higgs Doublet model} 
Non-zero EDMs (CEDMs) of quarks are the sign of CP violation  and they
are extremely small in the framework of the SM. This forces one to study 
the CP violating interactions in the new physics beyond the SM. In this 
section, we calculate top quark EDM and CEDM in the model III and take 
complex Yukawa couplings which are the possible sources of CP violation. 
The starting point is the general Yukawa interaction
\begin{eqnarray}
{\cal{L}}_{Y}=\eta^{U}_{ij} \bar{Q}_{i L} \tilde{\phi_{1}} U_{j R}+
\eta^{D}_{ij} \bar{Q}_{i L} \phi_{1} D_{j R}+
\xi^{U\,\dagger}_{ij} \bar{Q}_{i L} \tilde{\phi_{2}} U_{j R}+
\xi^{D}_{ij} \bar{Q}_{i L} \phi_{2} D_{j R} + h.c. \,\,\, ,
\label{lagrangian}
\end{eqnarray}
where $L$ and $R$ denote chiral projections $L(R)=1/2(1\mp \gamma_5)$,
$\phi_{i}$ for $i=1,2$, are the two scalar doublets, 
$\bar{Q}_{i L}$ are left handed quark doublets, $U_{j R} (D_{j R})$ are 
right handed up (down) quark singlets, with  family indices $i,j$. The 
Yukawa matrices $\eta^{U,D}_{ij}$ and $\xi^{U,D}_{ij}$ have in general 
complex entries. It is possible to collect SM particles in the first doublet 
and new particles in the second one by choosing the parametrization for 
$\phi_{1}$ and $\phi_{2}$ as \cite{Like, Atwood2}
\begin{eqnarray}
\phi_{1}=\frac{1}{\sqrt{2}}\left[\left(\begin{array}{c c} 
0\\v+H^{0}\end{array}\right)\; + \left(\begin{array}{c c} 
\sqrt{2} \chi^{+}\\ i \chi^{0}\end{array}\right) \right]\, ; 
\phi_{2}=\frac{1}{\sqrt{2}}\left(\begin{array}{c c} 
\sqrt{2} H^{+}\\ H_1+i H_2 \end{array}\right) \,\, .
\label{choice}
\end{eqnarray}
with the vacuum expectation values,  
\begin{eqnarray}
<\phi_{1}>=\frac{1}{\sqrt{2}}\left(\begin{array}{c c} 
0\\v\end{array}\right) \,  \, ; 
<\phi_{2}>=0 \,\, .
\label{choice2}
\end{eqnarray}
Here, $H_1$ and $H_2$ are the mass eigenstates $h^0$ and $A^0$ respectively 
since no mixing occurs between two CP-even neutral bosons $H^0$ and $h^0$ 
at tree level, for our choice. 

The part of the Yukawa lagrangian which is responsible for the Flavor 
Changing (FC) interaction is  
\begin{eqnarray}
{\cal{L}}_{Y,FC}=
\xi^{U\,\dagger}_{ij} \bar{Q}_{i L} \tilde{\phi_{2}} U_{j R}+
\xi^{D}_{ij} \bar{Q}_{i L} \phi_{2} D_{j R} + h.c. \,\, ,
\label{lagrangianFC}
\end{eqnarray}
where the couplings  $\xi^{U,D}$ for the FC charged interactions are
\begin{eqnarray}
\xi^{U}_{ch}&=& \xi^U_{N} \,\, V_{CKM} \nonumber \,\, ,\\
\xi^{D}_{ch}&=& V_{CKM} \,\, \xi^D_{N} \,\, ,
\label{ksi1} 
\end{eqnarray}
and $\xi^{U,D}_{N}$ is defined by the expression 
\begin{eqnarray}
\xi^{U (D)}_{N}=(V_{R (L)}^{U (D)})^{-1} \xi^{U,(D)} V_{L(R)}^{U (D)}\,\, .
\label{ksineut}
\end{eqnarray}
Notice that the index "N" in $\xi^{U,D}_{N}$ denotes the word "neutral". 

\begin{figure}[htb]
\vskip -3.0truein
\centering
\epsfxsize=6.8in
\leavevmode\epsffile{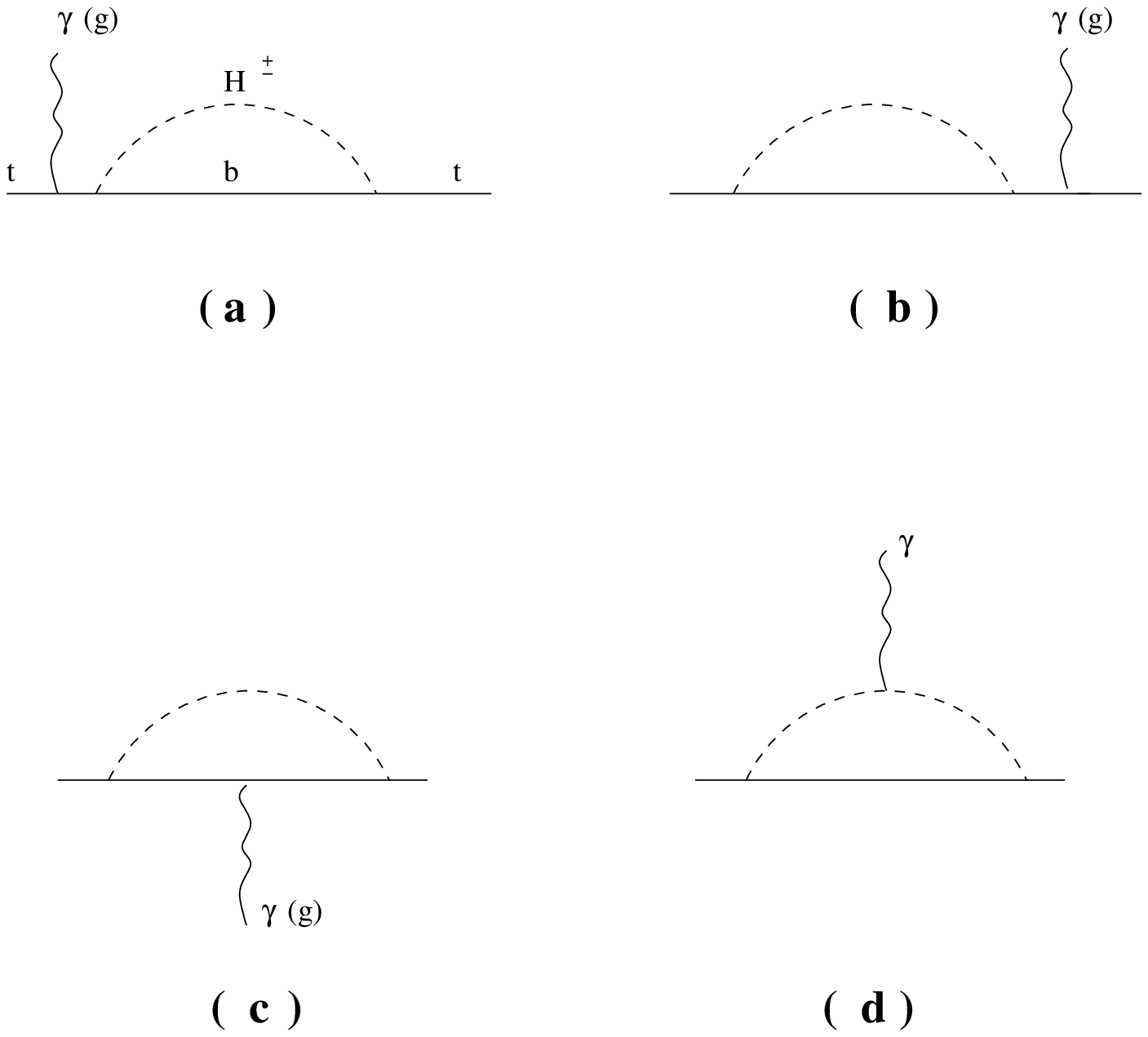}
\vskip -2.0truein
\caption[]{One loop diagrams contribute to EDM (CEDM) of top quark due to 
$H^{\pm}$ in the 2HDM. Wavy lines represent the electromagnetic
(chromomagnetic) field and dashed lines the $H^{\pm}$ field.}
\label{fig1}
\end{figure}

\begin{figure}[htb]
\vskip -3.0truein
\centering
\epsfxsize=6.8in
\leavevmode\epsffile{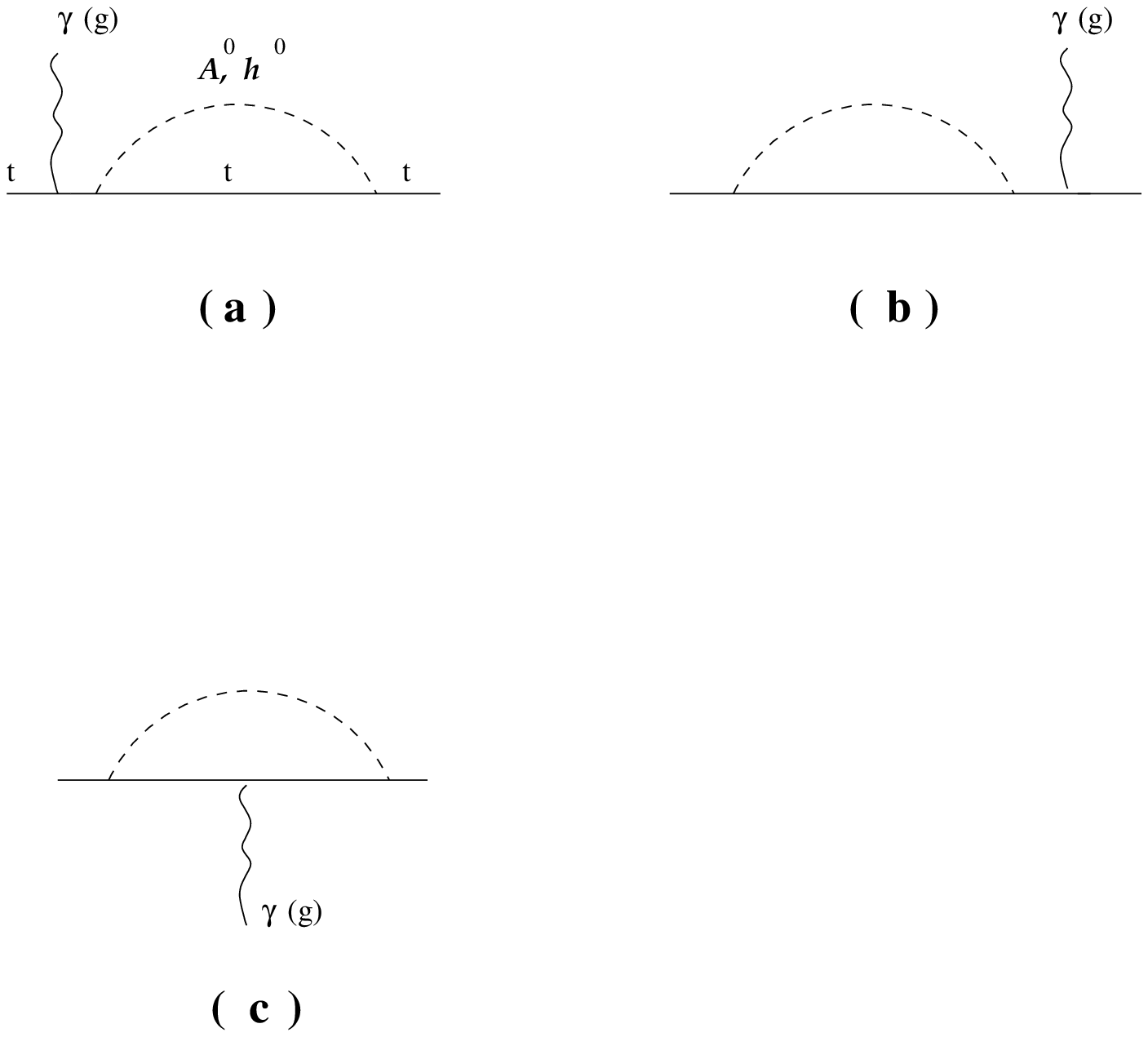}
\vskip -2.0truein
\caption[]{The same as Fig. \ref{fig1}, but for neutral Higgs bosons 
$h^0$ and $A^0$.}
\label{fig2}
\end{figure}

The effective EDM and CEDM interactions for $t$-quark are  
\begin{eqnarray}
{\cal L}_{EDM}=i e d_{\gamma} \,\bar{t}\,\gamma_5 \,\sigma^{\mu\nu}\,t\, 
F_{\mu\nu} \,\, ,
\label{EDM1}  
\end{eqnarray}
\begin{eqnarray}
{\cal L}_{CEDM}=i g_s d_g \,\bar{t}\,\gamma_5 \,\sigma^{\mu\nu}\, 
\frac{\lambda^{a}}{2}\,t\, G^a_{\mu\nu} \,\, ,
\label{CEDM1}  
\end{eqnarray}
where $F_{\mu\nu}$ and $G^a_{\mu\nu}$ are the electromagnetic and
choromodynamic field tensors, $\lambda^{a}$ are Gell Mann matrices with
color indices $a$, and "$d_{\gamma}$ ($d_g$)" is EDM (CEDM) of 
top quark and it is a real number by hermiticity. In the model III with 
complex Yukawa couplings, the charged $H^{\pm}$ and neutral Higgs bosons 
$h^0, A^0$ can induce CP violating interactions at loop level and this 
is the source of non-zero EDM and CEDM. We present the 1-loop diagrams 
due to charged and neutral Higgs particles in Figs. \ref{fig1} and \ref{fig2}. 
Notice that only vertex diagrams ($c$, $d$ in Fig. \ref{fig1} and $c$ in 
Fig. \ref{fig2}) contribute to the CP-violating interactions. 

The most general vertex operator for on-shell top quark and off-shell photon 
(gluon) can be written as
\begin{eqnarray}
\Gamma^{(a)}_{\mu}&=&F_1(q^2)\, \gamma_{\mu}\,(\frac{\lambda^{a}}{2})+ 
F_2 (q^2)\, \sigma_{\mu\nu}\,(\frac{\lambda^a}{2}) \,
q^{\nu}\nonumber \\ 
&+& F_3 (q^2)\, \sigma_{\mu\nu}\gamma_5\,(\frac{\lambda^{a}}{2})\, q^{\nu}
\label{vertexop}
\end{eqnarray}
where $q_{\nu}$ is photon (gluon) 4-vector and $q^2$ dependent form factors 
$F_{1}(q^2)$ and  $F_{2}(q^2)$ are proportional to the charge and 
anomalous (chromo) magnetic moment of top quark respectively. Existence of
the form factor $F_{3}(q^2)$ is the reason for the CP violating interactions 
and therefore for nonvanishing EDM (CEDM) of top quark. The EDM and CEDM of
top quark are obtained as a sum of contributions coming from charged and 
neutral Higgs bosons,
\begin{eqnarray}
d_{\gamma}=d_{\gamma}^{H^{\pm}}+d_{\gamma}^{h^0}+d_{\gamma}^{A^0}\,\,,
\label{EDM2}
\end{eqnarray}
and 
\begin{eqnarray}
d_{g}=d_{g}^{H^{\pm}}+d_{g}^{h^0}+d_{g}^{A^0}\,\, .
\label{CEDM2}
\end{eqnarray}
Using the equations 
\begin{eqnarray}
d^{H^{\pm}}&=&\frac{4\,G_F}{\sqrt {2}}\frac{1}{16\pi^2}\, 
\frac{m_b}{m_t^2}\,Im(\bar{\xi}^{D\,*}_{N,bb}\,\bar{\xi}^{U}_{N,tt})\, 
|V_{tb}|^2 \, \int_0^1 dx \,
\frac{(-1\,+\,x)\,\big(Q_b\, (-1\,+\,x)+ x\,\kappa \big )
\,y_t} {r_b\, y_t+x^2\, y_t-x\,(-1+y_t+r_b\,y_t)} 
\,\,,  \nonumber \\
d^{h^0}&=&\frac{4\,G_F}{\sqrt {2}}\frac{1}{16\pi^2}\, 
\frac{1}{m_t}\,Im(\bar{\xi}^{U}_{N,tt})\,Re(\bar{\xi}^{U}_{N,tt}) 
\, Q_t\, \Bigg\{ 1-\frac{r_1}{2}\,ln\,r_1 
\nonumber \\ &+& 
\frac{r_1\,(r_1-2)}{\sqrt{r_1\,(r_1-4)}}\Big (
Arctan\, \big( \frac{r_1}{\sqrt{r_1\,(r_1-4)}}\big ) -
Arctan\, \big(\frac{r_1-2}
{\sqrt{r_1\,(r_1-4)}} \big) \Big) \Bigg\}
\,\, , \mbox{ for $r_1<4$}, \nonumber \\
d^{h^0}&=&-\frac{4\,G_F}{\sqrt {2}}\frac{1}{16\pi^2}\, 
\frac{1}{m_t}\,Im(\bar{\xi}^{U}_{N,tt})\,Re(\bar{\xi}^{U}_{N,tt}) 
\, Q_t \, \Bigg\{ 1-\frac{r_1\,(r_1-2)}{\sqrt{r_1\,(r_1-4)}}\,
ln\frac{\sqrt{r_1}-\sqrt{r_1-4}}{2}
\nonumber \\ &-&
\frac{1}{2} r_1 \,ln\, r_1 \Bigg\} \,\, , \mbox{ for $r_1>4$}\, , 
\nonumber \\
d^{A^0}&=&-d_{\gamma}^{h^0} (r_1\rightarrow r_2)
\,\, ,
\label{EDM3}
\end{eqnarray}
we get $d_{\gamma \,(g)}^{H^{\pm}}$ , $d_{\gamma \,(g)}^{h^0}$ and 
$d_{\gamma \,(g)}^{A^0}$ as
\begin{eqnarray}
d_{\gamma}^{H^{\pm}}&=& d^{H^{\pm}}(\kappa =1) \,\, , \nonumber \\
d_{\gamma}^{h^0\,(A^0)}&=&d^{h^0\,(A^0)}\,\, ,
\label{EDM4}
\end{eqnarray}
and 
\begin{eqnarray}
d_g^{H^{\pm}}&=& d^{H^{\pm}}(\kappa =0, \, Q_b\rightarrow 1) \,\, , 
\nonumber \\
d_g^{h^0\,(A^0)}&=&d^{h^0\,(A^0)} (Q_t\rightarrow 1) \,\, .
\label{CEDM4}
\end{eqnarray}
Here $r_b=m_b^2/m_t^2$, $r_1=m_{h^0}^2/m_t^2$, $r_2=m_{A^0}^2/m_t^2$, 
$y_t=\frac{m_t^2}{m^2_{H^{\pm}}}$, $Q_b$ and $Q_t$ are charges of 
$b$ and $t$ quarks respectively and 
$\xi^{U(D)}_{N,ij}=\sqrt{\frac{4\, G_F}{\sqrt{2}}}\,
\bar{\xi}^{U(D)}_{N,ij}$. 

In eqs. (\ref{EDM3}), we take only internal $b\,(t)$-quark contribution for 
charged (neutral) Higgs interactions. Here, we assume that the Yukawa couplings 
$\bar{\xi}^{U}_{N,it},\, i=u,c $, and $\bar{\xi}^{D}_{N, bj},\, j=d,s $ are 
negligible compared to $\bar{\xi}^{U}_{N,tt}$ and $\bar{\xi}^{D}_{N,bb}$ 
(see \cite{eril2}). 

The Yukawa couplings $\bar{\xi}^{U}_{N,tt}$ and 
$\bar{\xi}^{D}_{N,bb}$ are complex in general and we take, 
\begin{eqnarray}
\bar{\xi}^{U}_{N,tt}=|\bar{\xi}^{U}_{N,tt}|\, e^{i \theta_{tt}}
\nonumber \, , \\
\bar{\xi}^{D}_{N,bb}=|\bar{\xi}^{D}_{N,bb}|\, e^{i \theta_{tb}} \, .
\label{xi}
\end{eqnarray}
With this parametrization, EDM and CEDM of top quark can be obtained as  
\begin{eqnarray}
d_{\gamma}&=&\frac{4\,G_F}{\sqrt {2}}\frac{1}{16\pi^2}\, 
\frac{1}{m_t}\,|\bar{\xi}^{D}_{N,bb}|\,|\bar{\xi}^{U}_{N,tt}|\, 
\Bigg\{ |V_{tb}|^2 \,\sqrt{r_b}\, sin\,(\theta_{tt}-\theta_{tb}) 
\int_0^1 dx \,\frac{(-1\,+\,x)\,\big( Q_b\, (-1\,+\,x)+ x \,\kappa \big )
\,y_t} {r_b\, y_t+x^2\, y_t-x\,(-1+y_t+r_b\,y_t)} 
\nonumber \\ &+& 
\frac{1}{2}\,sin\,2\,\theta_{tt}\, Q_t\, \Bigg( 
\frac{r_2}{2}\,ln\,r_2-\frac{r_1}{2}\,ln\,r_1 
\nonumber \\ &+& \frac{r_1\,(r_1-2)}{\sqrt{r_1\,(r_1-4)}}\Big (
Arctan\,\big( \frac{r_1}{\sqrt{r_1\,(r_1-4)}}\,\big)-Arctan\,
\big(\frac{r_1-2} {\sqrt{r_1\,(r_1-4)}} \big) \Big)
\nonumber \\ &-& 
\frac{r_2\,(r_2-2)}{\sqrt{r_2\,(r_2-4)}}\Big (
Arctan\,\big( \frac{r_2}{\sqrt{r_2\,(r_2-4)}}\,\big)-Arctan\,
\big(\frac{r_2-2} {\sqrt{r_2\,(r_2-4)}} \big) \Big)
\Bigg) \Bigg\} \,\, ,
\label{EDM5}
\end{eqnarray}
and 
\begin{eqnarray}
d_g&=& d_{\gamma} (\kappa =0, \, Q_b\rightarrow 1, \, Q_t\rightarrow 1) 
\,\, , 
\label{CEDM5}
\end{eqnarray}
for $r_1,\, r_2<4$. In the case of other possibilities for $r_1$ and $r_2$
the expression for $d_{\gamma}$ and $d_g$ can be obtained by using the eqs.
(\ref{EDM3}), (\ref{EDM4}) and (\ref{CEDM4}).  

Now, for the completeness, we present the $q^2$ dependencies of the form 
factors $d_{\gamma}(q^2)$ and $d_g (q^2)$. Here $q^2$ is the virtuality of
the outgoing photon and gluon, respectively. Using the functions 
\begin{eqnarray}
d^{H^{\pm}}(q^2)&=&-\frac{4\,G_F}{\sqrt {2}}\frac{1}{16\pi^2}\, |V_{tb}|^2 \,
\Big \{ m_b\,Im(\bar{\xi}^{D\,*}_{N,bb}\,\bar{\xi}^{U}_{N,tt})\, f_1-
i\,  \frac{m_t}{2}\,
(|\bar{\xi}^{D}_{N,bb}|^2-|\bar{\xi}^{U}_{N,tt}|^2)\,f_2 \Big \} \, ,
\nonumber \\
d^{h^0}(q^2)&=&\frac{4\,G_F}{\sqrt {2}}\frac{1}{4\pi^2\,m_t}\,
Im(\bar{\xi}^{U}_{N,tt})\,Re(\bar{\xi}^{U}_{N,tt}) \, Q_t\,f_3 (r_1)\, ,
\nonumber \\
d^{A^0}(q^2)&=&-d^{h^0} (q^2, r_1\rightarrow r_2)
\,\, ,
\label{EDM3q2}
\end{eqnarray}
and the equation
\begin{eqnarray}
d_{\gamma}(q^2) =d^{H^{\pm}}(q^2,\kappa \rightarrow 1)+
d^{h^0}(q^2) + d^{A^0}(q^2)\,\, ,
\label{CEEDM2}
\end{eqnarray}
we get 
\begin{eqnarray}
d_{\gamma}(q^2)&=&-\frac{4\,G_F}{\sqrt {2}}\, \frac{1}{16\pi^2} \,
\Bigg \{ |V_{tb}|^2 \, 
\Big ( m_b\, |\bar{\xi}^{D}_{N,bb}|\,|\bar{\xi}^{U}_{N,tt}| 
\, sin\,(\theta_{tt}-\theta_{tb})\,f_1 (\kappa \rightarrow 1) 
\nonumber \\ &-&
i\,  \frac{m_t}{2}\,(|\bar{\xi}^{D}_{N,bb}|^2-|\bar{\xi}^{U}_{N,tt}|^2)\,
f_2 (\kappa \rightarrow 1) \Big ) \,
\nonumber \\ &-&
\frac{2}{m_t}\, sin\,2\,\theta_{tt} \, Q_t\,(f_3 (r_1)-f_3 (r_2) \Bigg \} 
\,\, ,
\label{EDMtotq2}
\end{eqnarray}
and 
\begin{eqnarray}
d_{g}(q^2)&=& d_{\gamma}(q^2, \kappa\rightarrow 0, Q_t\rightarrow 1, 
Q_b\rightarrow 1)
\,\, .
\label{CEDMtotq2}
\end{eqnarray}
Here the functions $f_1$, $f_2$ and $f_3 (z)$ are  
\begin{eqnarray}
f_1&=& \int_{0}^{1} \int_{0}^{1-x} dx\,dy\, 
(\frac{Q_b\,(-1+x)}{\Delta}+\frac{\kappa\, x}{\Delta^{\prime}}) \, , 
\nonumber \\
f_2&=& \int_{0}^{1} \int_{0}^{1-x} dx\,dy\, 
x\,(-1+x+2\,y)\,(\frac{Q_b}{\Delta}+\frac{\kappa}{\Delta^{\prime}}) \, , 
\nonumber \\
f_3 (z) &=& \int_{0}^{1} dx \frac{-1+x}{\sqrt {r_q}} 
\frac{Arctan \Big( \frac{\sqrt {r_q}\,(-1+x)}
{\sqrt{-r_q\, (-1+x)^2+4\,(1+(-2+ z)\,x+x^2)}}\Big )}
{\sqrt{-r_q\, (-1+x)^2+4\,(1+(-2+ z)\,x+x^2)}} \,\, ,
\label{EDMfunc5}
\end{eqnarray}
with 
\begin{eqnarray}
\Delta &=& \frac{m_t^2}{y_t}\, \Big ( x^2\,y_t+ r_q\,y\,y_t\,(-1+y)+
x \big( 1+ y_t\,(r_q\,y+r_b-1) \big) \Big )
\,\, , \nonumber \\
\Delta^{\prime} &=& \frac{m_t^2}{y_t}\, \Big ( 1+x^2\,y_t+ r_q\,y\,y_t\,
(-1+y)+ x \big( -1+ y_t\,(r_q\,y+r_b-1) \big) \Big )
\,\, , \nonumber \\
r_q&=& \frac{q^2}{m_t^2} \,\, .
\label{Deltaprm}
\end{eqnarray}
%
\section{Discussion}
This section is devoted to the analysis of dependencies of EDM and CEDM of 
top quark on the CP violating parameters $sin\,\theta_{tb}$, 
$sin\,\theta_{tt}$, the masses of charged and neutral Higgs bosons. Notice
that EDM (CEDM) is obtained in the limit $q^2 \rightarrow 0$ 
(see eqs. (\ref{EDMtotq2}) and (\ref{CEDMtotq2})). Since there are large 
number of free parameters in the model III, such as Yukawa couplings, 
$\bar{\xi}^{U (D)}_{N, ij}$, the masses of new Higgs bosons, $H^{\pm}$, 
$h^0$ and $A^0$, there is a need to restrict them using the experimental 
measurements.  To find a constraint region for these free parameters we 
restrict the Wilson coefficient $C_7^{eff}$, which is the effective 
coefficient of the operator $O_7 = \frac{e}{16 \pi^2} \bar{s}_{\alpha} 
\sigma_{\mu \nu} (m_b R + m_s L) b_{\alpha} {\cal{F}}^{\mu \nu}$
(see \cite{alil1} and references therein), in the region 
$0.257 \leq |C_7^{eff}| \leq 0.439$. Here upper and lower limits were 
calculated using the CLEO measurement \cite{cleo2}
\begin{eqnarray}
Br (B\rightarrow X_s\gamma)= (3.15\pm 0.35\pm 0.32)\, 10^{-4} \,\, .
\label{br2}
\end{eqnarray}
and all possible uncertainities in the calculation of $C_7^{eff}$ 
\cite{alil1}. Using this restriction we get a constraint region for the
couplings $\bar{\xi}^{U}_{N, tt}$, $\bar{\xi}^{D}_{N, bb}$ and the CP 
violating parameters  $sin\,\theta_{tt}$ and $sin\,\theta_{tb}$. 
In our calculations we neglect all the Yukawa couplings except 
$\bar{\xi}^{U}_{N, tt}$ and $\bar{\xi}^{D}_{N, bb}$ since they are negligible 
due to their light flavor contents \cite{Cheng}. Furthermore, we also 
respect the constraint for the angle $\theta_{tt}$ and $\theta_{bb}$, due to 
the experimental upper limit of neutron electric dipole moment, 
$d_n<10^{-25}\hbox{e$\cdot$cm}$ (or the more recent results 
$|d_n|< 6.3\times 10^{-26}\hbox{e$\cdot$cm (90 \% C.L.)}$ \cite{Harrison}),
which leads to $\frac{1}{m_t m_b} Im(\bar{\xi}^{U}_{N, tt}\, 
\bar{\xi}^{* D}_{N, bb})< 1.0$ for $M_{H^\pm} \approx 200$ GeV \cite{david}. 
Notice that we take $h^0$ as the lighest Higgs boson and assume that the 
coupling $\bar{\xi}^{U}_{N, tt}$ has a small imaginary part. 

\begin{figure}[htb]
\vskip -3.0truein
\centering
\epsfxsize=6.8in
\leavevmode\epsffile{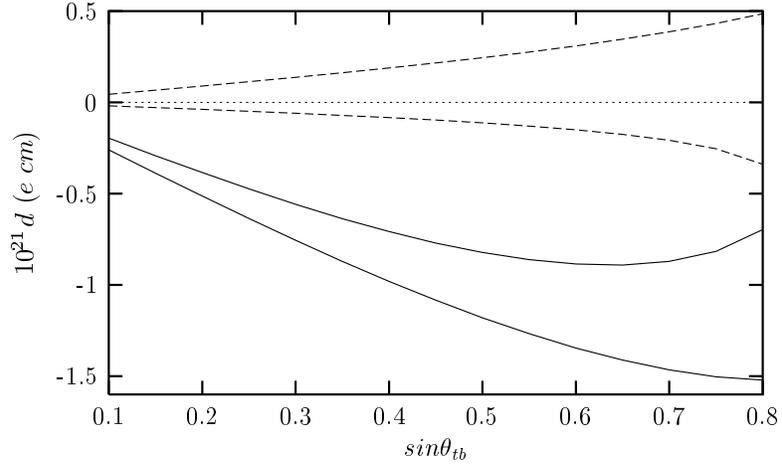}
\vskip -3.0truein
\caption[]{Top quark EDM "$d$" as a function of  $sin\,\theta_{tb}$ 
for $m_{H^{\pm}}=400\, GeV$, $sin\,\theta_{tt}=0$  and $|r_{tb}|<1$, in the
model III. Here $d$ is restricted in the region bounded by solid lines 
for $C_7^{eff}>0$ and by dashed  lines for $C_7^{eff}<0$}
\label{EDMchsintb}
\end{figure}

In  Fig. \ref{EDMchsintb}, we plot EDM "$d$" with respect to 
$sin\,\theta_{tb}$ for $m_{H^{\pm}}=400\, GeV$, 
$\bar{\xi}_{N, bb}^{D}=40\, m_b$ and 
$|r_{tb}|=|\frac{\bar{\xi}_{N, tt}^{U}}{\bar{\xi}_{N, bb}^{D}}| <1$, when 
the coupling $\bar{\xi}_{N, tt}^{U}$ is real. In this case the neutral 
Higgs bosons $h^0$ and $A^0$ do not have any contribution to the EDM 
of top quark. EDM is restricted in the region between solid (dashed) lines 
for $C_7^{eff} > 0$ ($C_7^{eff} < 0$). This physical quantity is at the 
order of the magnitude of $10^{-21}\,e\,cm$ for the large values of 
$sin\,\theta_{tb}$, especially for $C_7^{eff} > 0$. It can get both signs 
and even vanish for $C_7^{eff} < 0$. With the increasing values of 
$sin\,\theta_{tb}$, "$d$" becomes large as expected and the restricted region 
becomes wider, for both $C_7^{eff} > 0$ and $C_7^{eff} < 0$.

\begin{figure}[htb]
\vskip -3.0truein
\centering
\epsfxsize=6.8in
\leavevmode\epsffile{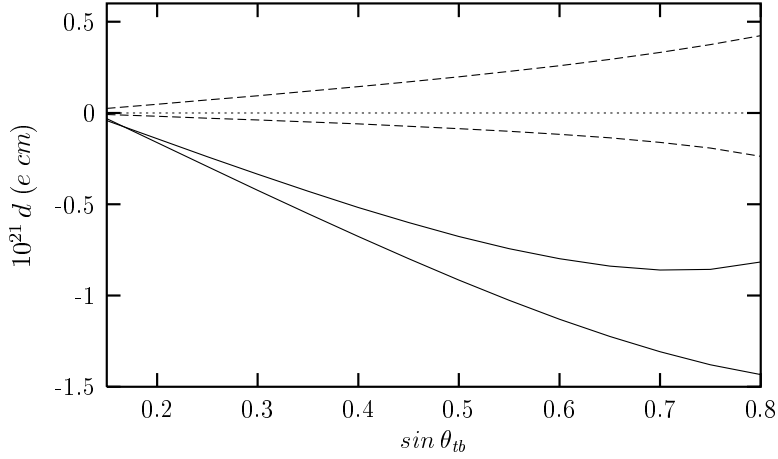}
\vskip -3.0truein
\caption[]{The same as Fig. \ref{EDMchsintb} but for $sin\,\theta_{tt}=0.1$, 
$m_{h^0}=80 \, GeV$ and $m_{A^0}=90\, GeV$}.
\label{EDMtotsintb}
\end{figure}

Fig. \ref{EDMtotsintb} is devoted to the EDM "$d$" with respect to 
$sin\,\theta_{tb}$ for $m_{H^{\pm}}=400\, GeV$, 
$\bar{\xi}_{N, bb}^{D}=40\, m_b$ and $|r_{tb}|<1$, when the coupling 
$\bar{\xi}_{N, tt}^{U}$ is complex. Here we take a small imaginary part 
for $\bar{\xi}_{N, tt}^{U}$, namely  $sin\,\theta_{tt}=0.1$. This is the 
case where the neutral Higgs bosons $h^0$ and $A^0$ have also contributions 
to the EDM of top quark. The magnitude of EDM slightly decreases for both 
$C_7^{eff} > 0$ and $C_7^{eff} < 0$ compared to the case where 
$\bar{\xi}_{N, tt}^{U}$ is real. 

\begin{figure}[htb]
\vskip -3.0truein
\centering
\epsfxsize=6.8in
\leavevmode\epsffile{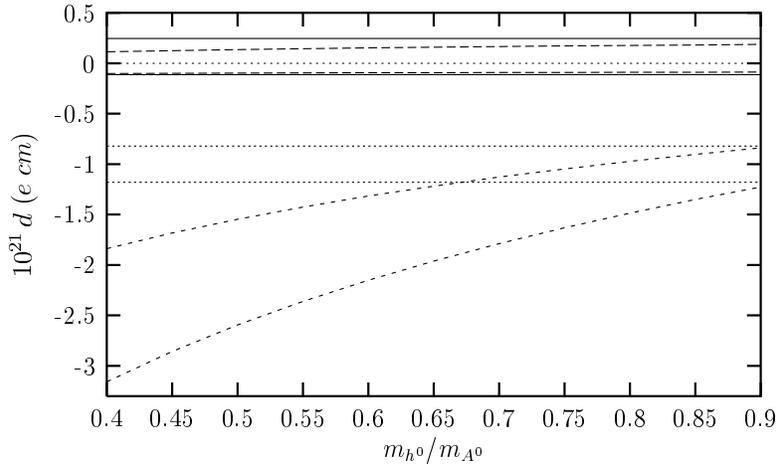}
\vskip -3.0truein
\caption[]{Top quark EDM "$d$" as a function of  the ratio $R_{neutr}=
\frac{m_{h^0}}{m_{A^0}}$, for $m_{H^{\pm}}=400\, GeV$, 
$sin\,\theta_{tb} = 0.5$, $sin\,\theta_{tt} = 0.1$,
$\bar{\xi}_{N,bb}^{D}=40\, m_b$ and $|r_{tb}|<1$, in the
model III. Here "$d$" lies in the region bounded by 
dashed (solid) lines for $C_7^{eff} < 0$ and by small dashed (dotted) lines 
for $C_7^{eff} > 0$, when neutral Higgs contributions are (not) taken into 
account.}
\label{EDMrat}
\end{figure}

In Fig. \ref{EDMrat} we present $R_{neutr}=\frac{m_{h^0}}{m_{A^0}}$
dependence of EDM for $sin\,\theta_{tb}=0.5$, $sin\,\theta_{tt}=0.1$, 
$m_{A^0}=90\, GeV$, $m_{H^{\pm}}=400\, GeV$, $\bar{\xi}_{N, bb}^{D}=40\, 
m_b$ and $|r_{tb}| <1$. Here "$d$" lies in the region bounded by 
solid lines for $C_7^{eff} < 0$ and by dotted lines for $C_7^{eff} > 0$, 
when the neutral Higgs contributions are not taken into account. With the
addition of the neutral Higgs boson contributions, "$d$" can reach to 
$3\times 10^{-21}\,(e\, cm)$ for $C_7^{eff} > 0$ and the small values of 
the ratio $R_{neutr}$. In the case of degenerate masses of $h^0$ and $A^0$,      
the neutral Higgs contributions vanish. 

For completeness we plot "$d$" with respect to the charged Higgs 
mass $m_{H^{\pm}}$ for $sin\,\theta_{tb}=0.5$, $sin\,\theta_{tt}=0.1$, 
$m_{h^0}=80\, GeV$, $m_{A^0}=90\, GeV$, $\bar{\xi}_{N,bb}^{D}=40\, m_b$, 
$C_7^{eff} > 0$ and a special choice of $|r_{tb}| <1$, $r_{tb}=0.001$. Here 
dashed (solid) line represents the charged (total) contribution. This 
figure shows that "$d$" is sensitive to the charged Higgs mass $m_{H^{\pm}}$.

CEDM is the EDM due to the external gluon instead of photon and it is also 
a clue about the existence of CP violating interactions in the theory.

In  Fig. \ref{CDMchsintb}, we plot CEDM "$d$" with respect to 
$sin\,\theta_{tb}$ for $m_{H^{\pm}}=400\, GeV$, $\bar{\xi}_{N, bb}^{D}=40\, m_b$, $|r_{tb}|<1$
and the real coupling $\bar{\xi}_{N, tt}^{U}$. CEDM is restricted in the region 
between solid (dashed) lines for $C_7^{eff} > 0$ ($C_7^{eff} < 0$). It can
reach to the order of magnitude of $\sim\, 10^{-20}\,g_s\,cm$ for large values 
of $sin\,\theta_{tb}$, for $C_7^{eff} > 0$. Similar to EDM, it can get both 
signs and even vanish for $C_7^{eff} < 0$. However, the sign of CEDM is
opposite to that of EDM. With the increasing values of $sin\,\theta_{tb}$, 
"$d$" becomes large as expected and the restricted region becomes wider, for 
both $C_7^{eff} > 0$ and $C_7^{eff} < 0$, similar to EDM.

Fig. \ref{CDMtotsintb} is devoted to the CEDM "$d$" with respect to 
$sin\,\theta_{tb}$ for $m_{H^{\pm}}=400\, GeV$, 
$\bar{\xi}_{N, bb}^{D}=40\, m_b$ and $|r_{tb}|<1$, when the coupling 
$\bar{\xi}_{N, tt}^{U}$ is complex, with a small imaginary part, 
$sin\,\theta_{tt}=0.1$. In this case, the neutral Higgs bosons $h^0$ and 
$A^0$ have also contributions to the CEDM of top quark and they decrease 
the magnitude of CEDM slightly for both $C_7^{eff} > 0$ and 
$C_7^{eff} < 0$  compared to the case where $\bar{\xi}_{N, tt}^{U}$ is 
real. 

In Fig. \ref{CDMrat}, we represent $R_{neutr}=\frac{m_{h^0}}{m_{A^0}}$
dependence of CEDM for $sin\,\theta_{tb}=0.5$, $sin\,\theta_{tt}=0.1$, 
$m_{A^0}=90\, GeV$, $m_{H^{\pm}}=400\, GeV$, $\bar{\xi}_{N, bb}^{D}=40\, 
m_b$ and $|r_{tb}| <1$. Here "$d$" lies in the region bounded by 
solid lines for $C_7^{eff} < 0$ and by dotted lines for $C_7^{eff} > 0$, 
when neutral Higgs contributions are not taken into account. A weak 
enhancement in the magnitude of CEDM appears with the addition of neutral 
Higgs boson contributions, for $C_7^{eff} > 0$ and small values of the
ratio $R_{neutr}$.

Finally, we plot CEDM with respect to the charged Higgs 
mass $m_{H^{\pm}}$ for $sin\,\theta_{tb}=0.5$, $sin\,\theta_{tt}=0.1$, 
$m_{h^0}=80\, GeV$, $m_{A^0}=90\, GeV$, $\bar{\xi}_{N,bb}^{D}=40\, m_b$, 
$C_7^{eff} > 0$ and $r_{tb}=0.001$. Here dashed (solid) line represents 
the charged (total) contribution. CEDM is sensitive to the charged Higgs 
mass $m_{H^{\pm}}$, similar to the EDM case.

Now we would like to summarize our results:

\begin{itemize}

\item EDM (CEDM) is generated by the one loop diagrams with the choice of 
complex Yukawa couplings in the model III. In the case of real 
$\bar{\xi}_{N,tt}^{U}$, only the charged Higgs sector contributes. The 
additional imaginary part of $\bar{\xi}_{N,tt}^{U}$ ensures that the neutral 
Higgs part can also have contribuion. In general, EDM and CEDM depends 
on the CP violating parameter $sin\,(\theta_{tt}-\theta_{tb})$ in the 
charged part and on $sin\,2\,\theta_{tt}$ in the neutral part.
  
\item Top quark EDM (CEDM) is sensitive to the ratio $R_{neutr}$, especially 
for $C_7^{eff} > 0$. The sensitivity to the charged Higgs mass $m_{H^{\pm}}$ 
is stronger. 

\item  If EDM (CEDM) is negative (positive), $C_{7}^{eff}$ can have both 
signs. However, if it is negative, $C_{7}^{eff}$ must be positive
(negative). This is observation is useful in the determination of the sign of 
$C_{7}^{eff}$.

\item EDM (CEDM) of top quark is at the order of the magnitude of 
$\sim 10^{-21} e \, cm$ ($\sim 10^{-20} g_s \, cm$). 

\end{itemize}

Therefore, the experimental investigations of the top quark EDM and CEDM 
give powerful informations about the physics beyond the SM.
\section{Acknowledgement}
This work was supported by Turkish Academy of Sciences (TUBA/GEBIP).

\newpage
\newpage

\newpage

\begin{figure}[htb]
\vskip -3.0truein
\centering
\epsfxsize=6.8in
\leavevmode\epsffile{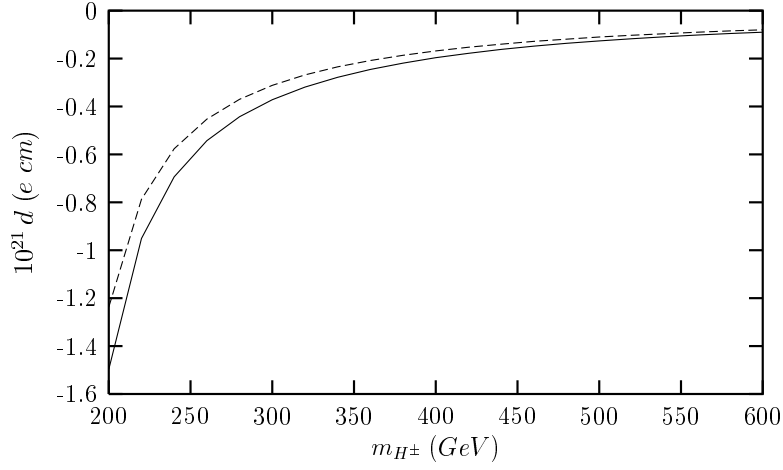}
\vskip -3.0truein
\caption[]{Top quark EDM "$d$" with respect to the charged Higgs 
mass $m_{H^{\pm}}$ for $sin\,\theta_{tb}=0.5$, $sin\,\theta_{tt}=0.1$, 
$m_{h^0}=80\, GeV$, $m_{A^0}=90\, GeV$, $\bar{\xi}_{N,bb}^{D}=40\, m_b$, 
$C_7^{eff} > 0$ and $r_{tb}=0.001$. Here dashed (solid) line represents 
EDM (not) including neutral Higgs contributions.}
\label{EDMmh}
\end{figure}
\begin{figure}[htb]
\vskip -3.0truein
\centering
\epsfxsize=6.8in
\leavevmode\epsffile{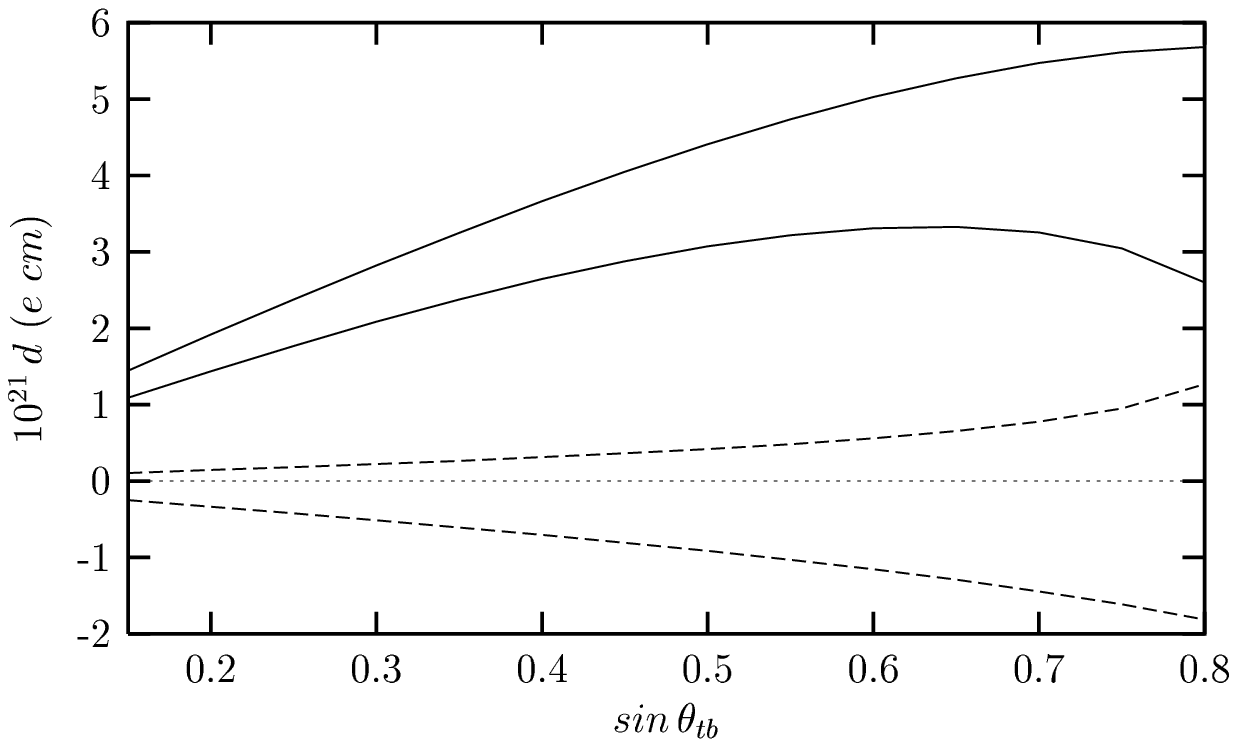}
\vskip -3.0truein
\caption[]{The same as Fig. \ref{EDMchsintb}, but in for top quark CEDM.}
\label{CDMchsintb}
\end{figure}
\begin{figure}[htb]
\vskip -3.0truein
\centering
\epsfxsize=6.8in
\leavevmode\epsffile{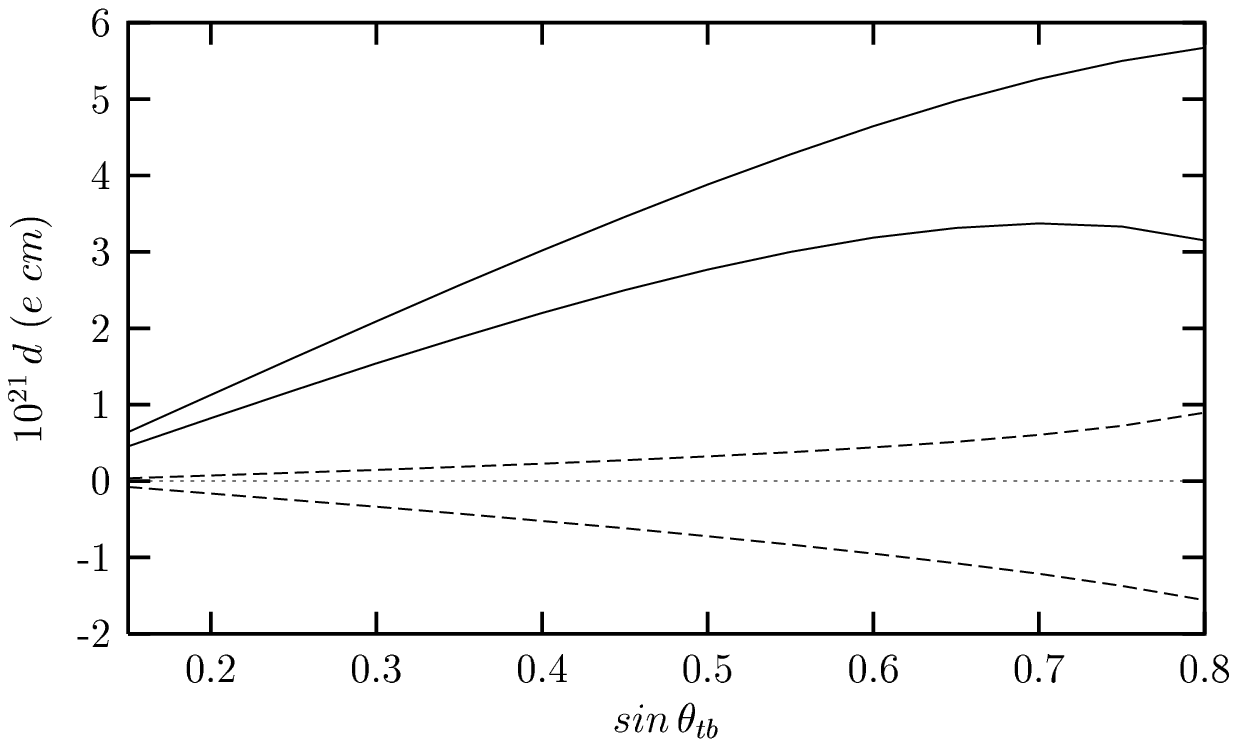}
\vskip -3.0truein
\caption[]{The same as Fig. \ref{EDMtotsintb}, but in for top quark CEDM.}
\label{CDMtotsintb}
\end{figure}
\begin{figure}[htb]
\vskip -3.0truein
\centering
\epsfxsize=6.8in
\leavevmode\epsffile{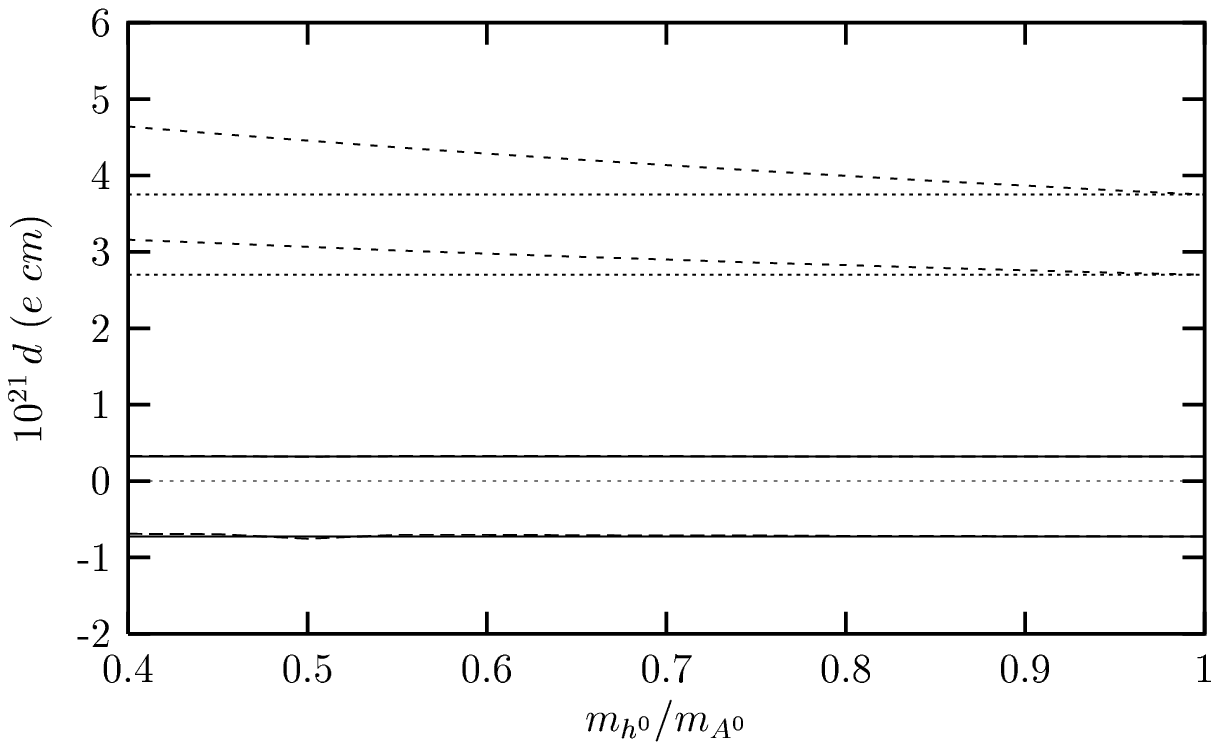}
\vskip -3.0truein
\caption[]{The same as Fig. \ref{EDMrat}, but in for top quark CEDM.}
\label{CDMrat}
\end{figure}
\begin{figure}[htb]
\vskip -3.0truein
\centering
\epsfxsize=6.8in
\leavevmode\epsffile{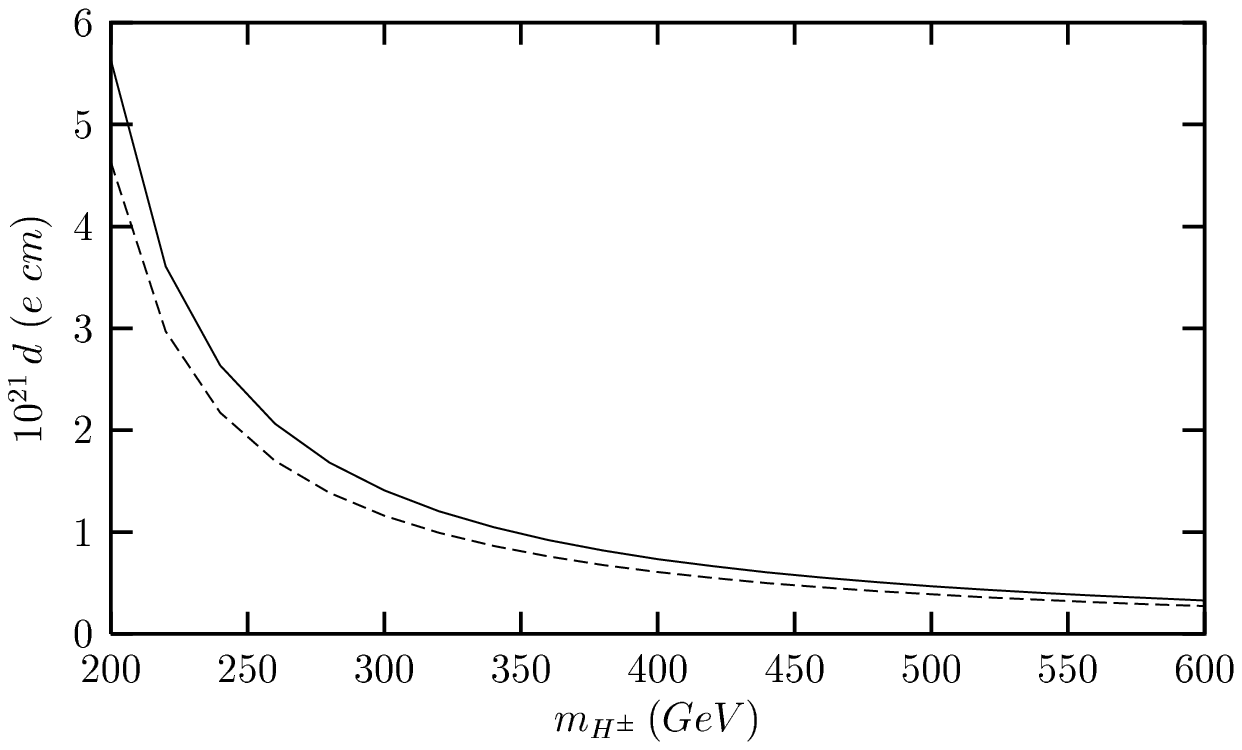}
\vskip -3.0truein
\caption[]{The same as Fig. \ref{EDMmh}, but in for top quark CEDM.}
\label{CDMchsintetb}
\end{figure}

\begin{thebibliography}{1}
%
\bibitem{smith} K. F. Smith et.al, {\it Phys. Lett.} {\bf B234} (1990) 191,
2885; I. S. Altarev et. al, {\it Phys. Lett.} {\bf B276} (1992) 242.
%
\bibitem{Commins} E. D. Commins et.al. {\it Phys. Rev. A} {\bf 50} 
(1994) 2960.  
%
\bibitem{Bailey} J. Bailey et al, {\it Journ. Phys.} {\bf G4} 
(1978) 345; 
%
\bibitem{Groom} Particle Data Group, D. E. Groom et.al. {\it Eur.    
Phys. J.} {\bf C15} (2000) 1. 
%
\bibitem{Khiplovich} I. B. Khriplovich,  {\it Yad. Fiz.} 
{\bf 44} (1986) 1019 ({\it Sov. J. Nucl. Phys.} {\bf 44} (1986) 659);  
{\it Phys. Lett.} {\bf B173} (1986) 193;
A. Czarnecki and B. Krause, {\it Acta Phys. Polon.} {\bf B28} (1997) 829. 
%
\bibitem{sahab1} E. P. Shabalin, {\it Sov. J. Nucl. Phys.} {\bf 28} (1978) 75. 
%
\bibitem{donog} J. F. Donoghue, {\it Phy. Rev.} {\bf D18} (1978) 1632. 
%
\bibitem{krause} A. Czarnecki and B. Krause, {\it Phys. Rev. Lett.}
{\bf 78} (1997) 4339.
%
\bibitem{Schmidt} C. R. Schmidt and M. E. Peskin, {\it Phys. Rev. Lett.}
{\bf 69} (1992) 410.
%
\bibitem{Weinberg} S, Weinberg, {\it Phys. Rev. Lett.}
{\bf 36 } (1976) 294; N. G. Deshpande and E. Ma, {\it Phys. Rev.}{\bf
D 16 } (1977) 1583; for a recent review, see H. Y. Cheng, {\it Int. J. Mod.
Phys.}{\bf A 7} (1992) 1059.  
%
\bibitem{Branco} G. C. Branco and M. N. Rebelo,{\it Phys. Rev. Lett.}
{\bf B 160 } (1985) 117; J. Liu and L. Wolfenstein, {\it Nucl. Phys..}{\bf 
B 289 } (1987) 1. 
%
\bibitem{Smith} C. H. Albright, J. Smith and S. H. H. Tye, {\it Phys. Rev.}
{\bf D 21 } (1980) 711.
%
\bibitem{Atwood1} D. Atwood. et. al., {\it Phys. Rev.} {\bf D 51 } (1995) 1034.                                          
%
\bibitem{liao} Y. Liao and X. Li, {\it Phys. Rev.}
{\bf D60 } (1999) 073004.
%
\bibitem{Dumm} D. G. Dumm and G. A. G Sprinberg, {\it Eur. Phys. J.}
{\bf C 11 } (1999) 293.
%
\bibitem{WSHou} W. S. Hou, {\it Phys. Lett.} {\bf B 296} (1992) 179, 
D. Chang, W. S. Hou and W. Y. Keung, {\it Phys. Rev.} {\bf D 48 } (1993) 
217.                                          
%
\bibitem{eril3} E. Iltan, {\it J. Phys.} {\bf G27} (2001) 1723.
%
\bibitem{Xu} A. Soni and R. M. Xu, {\it Phys. Rev. Lett.}  
{\bf 69 } (1992) 33.                                          
%
\bibitem{Like} M. E. Luke and M. J. Savage, {\it Phys. Lett.} {\bf B 307} 
(1993) 387.
%
\bibitem{Atwood2} D. Atwood et. al., {\it Phys. Rev.} {\bf D 54} 
(1996) 3296. 
%
\bibitem{eril2} E. Iltan, {\it Phys. Rev.} {\bf D60} (1999) 034023.
%
\bibitem{alil1} T. M.Aliev, E. Iltan, {\it J. Phys.} {\bf G25}
(1999) 989.
%
\bibitem{cleo2} M. S. Alam Collaboration, to appear in ICHEP98 Conference
(1998)
%
\bibitem{Cheng} T. P. Cheng and M. Sher, {\it Phys. Rev.} {\bf D 35} (1987) 
3484, M. Sher and Y. Yuan, {\it Phys. Rev.} {\bf D 44} (1991) 1461
%
\bibitem{Harrison} P. G. Harris et.al., {\it Phys. Rev. Lett.} {\bf 82} (1999) 
904.
%
\bibitem{david} D. B. Chao, K. Cheung and W. Y. Keung,
{\it Phys. Rev.} {\bf 59} (1999) 115006.

%
\end{thebibliography}
\end{document}